\documentclass[twocolumn,lengthcheck,superscriptaddress]{revtex4-1}

\usepackage{lmodern}
\usepackage{amsmath}
\usepackage{amsthm}
\usepackage{amssymb}
\usepackage{amsfonts}
\usepackage{graphicx}
\usepackage{hyperref}
\usepackage{color}
\usepackage{braket}

\newcommand{\tr}{\operatorname{tr}}

\def\su2{\textsl{SU}(2)}
\def\o3{\textsl{O}(3)}

\def\u2{\textsl{U}(2)}

\def\gl2c{\textsl{GL}(2, \mathbb{C})}
\def\lasu2{\mathfrak{su}(2)}

\theoremstyle{definition}

\theoremstyle{remark}

\begin{document}

\title{Holographic fluctuations and the principle of minimal complexity}

\author{Wissam Chemissany}
\affiliation{Institut f\"ur Theoretische Physik, Leibniz Universit\"at Hannover, Appelstr. 2, 30167 Hannover, Germany}
\email{wissamch@mit.edu,\,wissam.chemissany@itp.uni-hannover.de,\,\,tobias.osborne@itp.uni-hannover.de }
\affiliation{Research Laboratory of Electronics, Massachusetts Institute of Technology, Cambridge, Massachusetts 02139, USA}

\author{Tobias J.\ Osborne}
\affiliation{Institut f\"ur Theoretische Physik, Leibniz Universit\"at Hannover, Appelstr. 2, 30167 Hannover, Germany}

\begin{abstract}
We discuss, from a quantum information perspective, recent proposals of Maldacena, Ryu, Takayanagi, van Raamsdonk, Swingle, and Susskind that spacetime is an emergent property of the quantum entanglement of an associated boundary quantum system. We review the idea that the informational principle of minimal complexity determines a dual holographic bulk spacetime from a minimal quantum circuit $U$ preparing a given boundary state from a trivial reference state. We describe how this idea may be extended to determine the relationship between the \emph{fluctuations} of the bulk holographic geometry and the fluctuations of the boundary low-energy subspace. In this way we obtain, for every quantum system, an Einstein-like equation of motion for what might be interpreted as a bulk gravity theory dual to the boundary system.
\end{abstract}

\maketitle

\section{Introduction}
Holographic duality is the fascinating proposal that quantum field theories of a \emph{boundary} system are \emph{dual} to quantum gravity theories of an associated higher-dimensional \emph{bulk} spacetime. This proposal found a stunningly precise realisation in the work of Maldacena \cite{Maldacena:1998a, Maldacena:1999a} who argued that there is an exact equivalence between string theory on $\text{AdS}_5\times S^5$ and $\mathcal{N}=4$ supersymmetric Yang-Mills theory on the four-dimensional boundary. This was quickly solidified by Gubser, Klebanov, and Polyakov \cite{Gubser:1998a} and Witten \cite{Witten:1998a}. Since these foundational works there has been a huge amount of effort exporing such AdS/CFT dualities. Most recently, quantum information ideas have been exploited to provide microscopic toy models to understand quantum gravity \cite{Lloyd:2005a} and bulk/boundary correspondences \cite{Pastawski:2015a, Yang:2016a, Yang:2016b}.

The idea that a bulk holographic spacetime might be associated with the entanglement structure of a boundary quantum system finds its antecedents in the early works of Jacobsen \cite{Jacobsen:1995a} and Holzhey, Larsen, and Wilczek \cite{Holzhey:1994a}: Jacobsen argued that Einstein's equations arise from black hole thermodynamics and might find their best  interpretation as an equation of state (see \cite{Bianchi:2012ev} for a thorough account and references). By combining Jacobsen's observation with the earlier derivations of the area law of entanglement in conformal field theory \cite{Holzhey:1994a} one could already see a kernel of later developments in embryonic form. 

The precise connection between bulk geometries and the structure of entanglement of low-energy states of a boundary system was realised by Ryu and Takayanagi, who conjectured --- based on analogies with black hole entropy via the AdS/CFT correspondence --- that the amount of entanglement on the boundary of the spacetime is given by the area (in Planck units) of certain extremal surfaces (of co-dimension $2$) in the bulk  \cite{Ryu:2006bv}. The Ryu-Takayanagi conjecture was later reduced to the original AdS/CFT relation by Lewkowycz and Maldacena \cite{Lewkowycz:2013nqa}. However, it took until Van Raamsdonk's essay \cite{VanRaamsdonk:2010pw} before the full scale of the connection between quantum entanglement, as geometric glue,  and quantum gravity began to be emerge. During the same year, Swingle had independently drawn in \cite{Swingle:2009bg} largely the same conclusion as Van Raamsdonk. Further arguments for the connection between entanglement and geometry via tensor networks were then developed in \cite{Evenbly:2011a}. Swingle and Van Raamsdonk later coauthored an investigation into dynamics: they have since managed to derive Einstein's equations linearized around pure AdS \cite{Swingle:2014uza}, providing further evidence that the dynamics of spacetime, as well as its geometry, indeed emerge from the structure of entanglement. Concurrently, Maldacena and Susskind \cite{Maldacena:2013xja} put forward their ER=EPR conjecture according to which a wormhole is equivalent to an entangled pair of black holes--significantly strengthening support for the idea of geometrising entanglement.

The proposals we discuss are found in recent works \cite{Susskind:2014moa,Stanford:2014jda,Brown:2015bva,Brown:2015bva2}  and talks 
\cite{Swingletalk2015, Susskindtalk2015,VanRaamsdonktalk2015} of van Raamsdonk, Swingle, Susskind, and Stanford: the core idea we explore is that the pattern of the entanglement of a (boundary) state $|\psi\rangle$ of a collection of degrees of freedom (qubits for simplicity) determines a dual bulk holographic spacetime via the \emph{principle of minimal complexity}. In particular, in this paper we discuss a precise approach to associating a bulk geometry, as a \emph{topological space}, with a quantum system comprised of a discrete collection of degrees of freedom and discuss the relationship between fluctuations of the bulk geometry and perturbations of the boundary quantum system. To that end, in the next section we review the prerequisite material and introduce all the necessary preliminary machinery to discuss correlated quantum systems and bulk geometries. In Sec.~\ref{sec:btg} we introduce two alternative ways, both capturing the essence of the principle of minimal complexity, to associate a bulk holographic spacetime, as a topological space, with the low-energy sector of a strongly correlated boundary quantum system. Following this, in Sec.~\ref{sec:cabf} we introduce an action, building on the principle of minimal complexity, to model fluctuations of the bulk holographic spacetime. The connection between boundary perturbations and bulk fluctuations is then developed in Sec.~\ref{sec:bpjf}, where Jacobi fields play a prominent role. These ideas are then explored in the context of several simple examples in Sec.~\ref{sec:examples}. Finally, in Sec.~\ref{sec:conclusions} we present our conclusions and outlook. 

\section{Preliminaries}
The language and notation we use throughout this paper is influenced by that employed in the literature on the AdS/CFT correspondence; we summarise it here briefly to orient the reader. Firstly, we refer throughout to two rather different systems, namely, the \emph{bulk} $\mathcal{M}$ and the \emph{boundary} $\partial \mathcal{M}$. In the AdS/CFT context the bulk system $\mathcal{M}$ is the AdS spacetime and the boundary $\partial\mathcal{M}$ is the CFT. Here the boundary system $\partial \mathcal{M}$ is taken to be a quantum system comprised of $n$ distinguishable subsystems. One particular example plays a prominent role throughout this paper, namely that of $n$ \emph{qubits} where $\partial \mathcal{M}$ has Hilbert space given by $\mathcal{H} \equiv \bigotimes_{j=1}^n \mathbb{C}^2$. (The calculations for the qubit case are representative of more complicated examples such as qudits or even harmonic oscillators, in which case the boundary Hilbert space is given by $\mathcal{H} \equiv \bigotimes_{j=1}^n L^2(\mathbb{R})$.) The bulk system is a ``\emph{classical system}'' which, for the purposes of this paper, is taken to be a \emph{topological space} $(X, \mathcal{T})$ with point set $X \cong \{1,2, \ldots, n\}\times \mathbb{R}^{+}$ and an, as yet undetermined, topology $\mathcal{T}$. The point set $X$ corresponds to a partially discretised \emph{holographic spacetime} with discrete boundary ``spatial'' coordinates and an additional continuous ``holographic time'' or ``radial'' coordinate referred to, henceforth, as $r\in\mathbb{R}^{+}$. Since the boundary system is a standard quantum system, and we are working in the Hamiltonian picture, there is an additional ``standard time coordinate'' $\tau$ (corresponding to the usual time for a boundary CFT); we always work on a single time slice for both the boundary and bulk and hence this coordinate is suppressed throughout. Thus, unless otherwise specified, whenever we say ``time $r$'' we are referring to the holographic time/radial coordinate. 

The boundary system is intended to capture \emph{all} of the \emph{relevant} low-energy degrees of freedom of some \emph{boundary Hamiltonian} $H\in\mathcal{B}(\mathcal{H})$. For example, if $H\ge 0$ is \emph{gapped} with a unique ground state then there is only \emph{one} relevant low-energy degree of freedom, namely the ground state $|\Omega\rangle$, in which case the boundary Hilbert space is just $\mathcal{H}\cong \mathbb{C}$. A slightly more nontrivial example is that of a ferromagnet in a small magnetic field where the relevant degrees of freedom are the vacuum and the single-magnon sector; here the relevant Hilbert space is $\mathcal{H} \cong \mathbb{C}^{n+1}$. A somewhat nontrivial example is that of the Hubbard model with $n$ sites at half filling with large on-site repulsion, in which case only the spin degrees of freedom are relevant and thus $\mathcal{H} \cong \bigotimes_{j=1}^n \mathbb{C}^2$. A final example, which we don't pursue here, is that of a system of $n$ anyons in general position. In this case $\text{dim}(\mathcal{H}) \propto d^n$, where $d$ is the total quantum dimension. 

The boundary Hamiltonians $H$ are taken to be \emph{local} with respect to some finite simple graph $G \equiv (V,E)$, where $V$ is the \emph{vertex set} representing the $n$ subsystems and $E$ is the \emph{edge set} representing interactions, i.e.,
\begin{equation}
	H = \sum_{j\sim k} h_{jk},
\end{equation}
where $h_{jk}$ are hermitian operators acting nontrivially only on subsystems $j$ and $k$ and as the identity otherwise, and $j\sim k$ means that $(j,k)$ is an edge of the graph $G$.

States of the boundary Hilbert space $\mathcal{H}$ may be specified in terms of a trivial reference basis, henceforth called the \emph{computational basis}, which is usually determined by a \emph{trivial} or \emph{elementary} initial local Hamiltonian. For our quantum spin system this is just the product basis $|x_1x_2\cdots x_n\rangle$, $x_j\in \{0,1\}$, $j = 1, 2, \ldots, n$ (for a system of harmonic oscillators, this would be the overcomplete basis $|\alpha_1\alpha_2\cdots \alpha_n\rangle$, $\alpha_j\in \mathbb{C}$, $j = 1, 2, \ldots, n$, of all coherent states). The boundary Hamiltonian determines a second basis via the unitary $U$ which diagonalises $H$, i.e., $U^\dag HU = D$, with $D$ diagonal. Because global phases are irrelevant the unitary $U$ may be understood as an element of the \emph{special unitary group} $\textsl{SU}(\mathcal{H}) \cong \textsl{SU}(2^n)$. It is worth noting that even if $H$ is rather simple, e.g., $G$ is a line graph, that $U$ can be extremely difficult to determine in general (see, e.g., 
\cite{Osborne:2011a, Aharonov:2013a, Gharibian:2015a} and references therein for examples).  

The unitary $U$ diagonalising the boundary Hamiltonian $H$ is the central object of interest here: its entangling structure determines an associated dual holographic bulk spacetime $\mathcal{M}$. The way this is done is by studying the \emph{quantum information complexity} of $U$ counting the number of nontrivial quantum gates required to implement $U$. A powerful method to precisely capture the information complexity of a unitary $U\in\textsl{SU}(\mathcal{H})$ was introduced by Nielsen and coauthors \cite{Nielsen:2005a, Nielsen:2006a, Nielsen:2006b, Dowling:2007a, Drezgich2007a, Shizume:2012a}, who proposed, for certain specific metrics on the tangent space $T_{U} \textsl{SU}(\mathcal{H})$ of $\textsl{SU}(\mathcal{H})$ at $U$,
$$\langle\cdot,\cdot\rangle_U : T_{U} \textsl{SU}(\mathcal{H}) \times T_{U} \textsl{SU}(\mathcal{H}) \rightarrow \mathbb{R},$$
the \emph{geodesic length} ${C}(U) \equiv d(\mathbb{I}, U)$ between the identity $\mathbb{I} \in \textsl{SU}(\mathcal{H})$ and $U$ as an appropriate measure, where
\begin{equation}\label{eq:geodesicdist}
	d(\mathbb{I},U) \equiv \inf_{\gamma}\int \sqrt{\langle K(r), K(r) \rangle}\, dr,
\end{equation}
and the infimum is over all curves $\gamma(r)\in \textsl{SU}(\mathcal{H})$ with tangent vector $-iK(r)\gamma(r)$ connecting $U$ to the identity $\mathbb{I}$, i.e., we have, via integration of the Schr\"odinger equation $\partial_r \gamma(r) =-iK(r) \gamma(r)$, that $\gamma(0) = \mathbb{I}$ and $\gamma(R) = U$, for some $R\in\mathbb{R}^{+}$. 

All the metrics in this paper are taken to be right invariant by identifying the tangent space at $\mathbb{I}$ with that at $U \in \textsl{SU}(\mathcal{H})$ via $-iK\mapsto -iKU$, where $-iK\in\mathfrak{su}(\mathcal{H})$ is a tangent vector \footnote{Tangent vectors $K\in\mathfrak{su}(\mathcal{H})$ are hence antihermitian operators of the form $K = -ik$, with $k\in\mathcal{B}(\mathcal{H})$ hermitian.} at $\mathbb{I} \in \textsl{SU}(\mathcal{H})$. Accordingly the metric $\langle\cdot,\cdot\rangle_U$ is constant as a function of $U$ and we henceforth write $\langle\cdot,\cdot\rangle_U \equiv \langle\cdot,\cdot\rangle$. One particular family of metrics plays a key role in this paper, namely
\begin{equation}
	\langle A,B\rangle_p \equiv \frac{1}{\dim(2^n)}\tr(\mathcal{D}_p^{\otimes n}(A^\dag)\mathcal{D}_p^{\otimes n}(B)),
\end{equation}
where 
\begin{equation}
	\mathcal{D}_p(X) = (1-p)\tr(X)\frac{\mathbb{I}}{2} + p X,
\end{equation}
with $p\in \mathbb{R}^+$. When $p\in [0,1]$ this is the \emph{depolarising channel}. For the special case that $p=1$ this metric reduces to the standard right-invariant metric on $\textsl{SU}(\mathcal{H})$:
\begin{equation}
	\langle A,B\rangle \equiv \frac{1}{\dim(\mathcal{H})}\tr(A^\dag B).
\end{equation} 
In general, as $p\rightarrow \infty$ is increased, the measure $d(\mathbb{I},U)$ admits the pleasing operational interpretation as (being proportional to) the minimal number of quantum gates required to (approximately) implement $U$ as a quantum circuit \cite{Nielsen:2006a, Nielsen:2006b, Dowling:2007a, Drezgich2007a, Shizume:2012a}. The case $p=1$ does not admit as natural an operational interpretation as the $p\gg1$ case, nevertheless, we carry out most of our example calculations with respect to the $p=1$ metric because it so much easier. (Note, however, all the conclusions we draw in this paper hold also for the general case $p\in\mathbb{R}^{+}$.)

The metrics $\langle \cdot, \cdot\rangle_p$ are all examples of right-invariant metrics on a Lie group. This class of metric allows for elegant computations; the vector field $-iK(r)$ associated with the geodesic flow $\gamma(r)$ satisfies a compact equation known as the \emph{Euler-Arnol'd equation}
\begin{equation}
	-i\frac{dK(r)}{dr} = B_p(-iK(r),-iK(r)),
\end{equation}
where $B_p(\cdot,\cdot)$ is a bilinear form determined by $\langle [X,Y],Z\rangle_p \equiv \langle B(Z, Y), X\rangle_p$, $\forall X,Y,Z\in \mathfrak{su}(\mathcal{H})$ \cite{Arnold:1966a, Arnold:1989a, Wald:1984a}. In the special case $p=1$ and when $U$ is sufficiently close to $\mathbb{I}$, i.e., $\mathbb{I}$ and $U$ are not \emph{conjugate points} of $\textsl{SU}(\mathcal{H})$, then the geodesic $\gamma(r)$ is simply given by
\begin{equation}
	\gamma(r) \equiv e^{-irK},
\end{equation}
where $K\equiv i\log(U)$ is \emph{constant}.

The Nielsen complexity measure was taken up by Susskind and coworkers as a central tool to determine a bulk holographic space $\mathcal{M}$ from a \emph{state} $|\psi\rangle$ of the boundary space $\partial \mathcal{M}$ specified by $H$. Here the idea is as follows. Take as input a quantum state $|\psi\rangle\in\mathcal{H}$ of the boundary Hilbert space and first find the unitary $U$ of \emph{minimal complexity} $C(U)$ which prepares $|\psi\rangle$ from an initial trivial state $|00\cdots 0\rangle$, i.e., $U|00\cdots 0\rangle = |\psi\rangle$. Now, assuming that the infimum in Eq.~(\ref{eq:geodesicdist}) may be \emph{achieved} by the geodesic $\gamma(r)$ with tangent vector $-iK(r)$, we can write 
\begin{equation}
	U \equiv \mathcal{T}e^{-i\int_0^R K(r)\, dr},
\end{equation}
where $\mathcal{T}$ denotes time ordering. This expression may then be approximated by discretisation: we find a \emph{quantum circuit} $V \equiv V_TV_{T-1}\cdots V_1$, where $V_j$, $j=1, 2, \ldots, T$, are \emph{quantum gates} acting on one or two qubits at a time, such that $V\approx U$:

\includegraphics{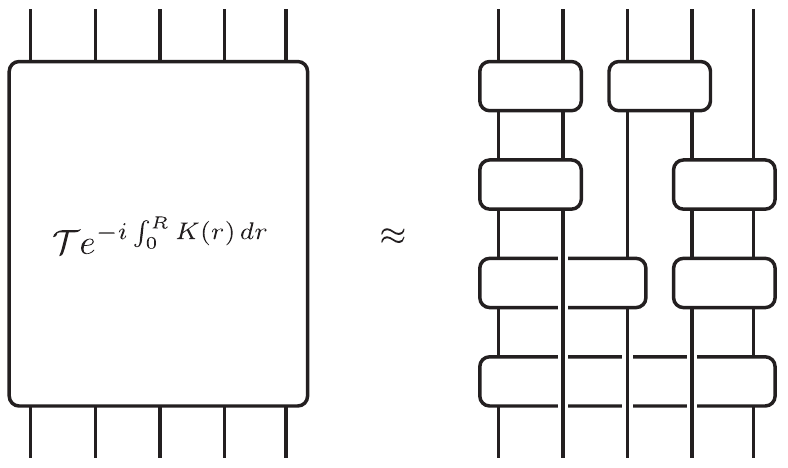}

That this can always be done is not totally trivial; see \cite{Berry:2015a, Berry:2015b} for the state of the art. The \emph{spacetime history} of the circuit $V$ determines a connectivity or adjacency relation on the \emph{vertex} or \emph{point set} $X \equiv \{1,2, \ldots, n\}\times \{1,2, \ldots, T\}$: we place an edge between vertices $(j,t) \in X$ and $(k,t) \in X$ if the two-qubit gate $V_t$, $t\in \{1,2, \ldots, T\}$, acts nontrivially on qubits $j$ and $k$:

\includegraphics{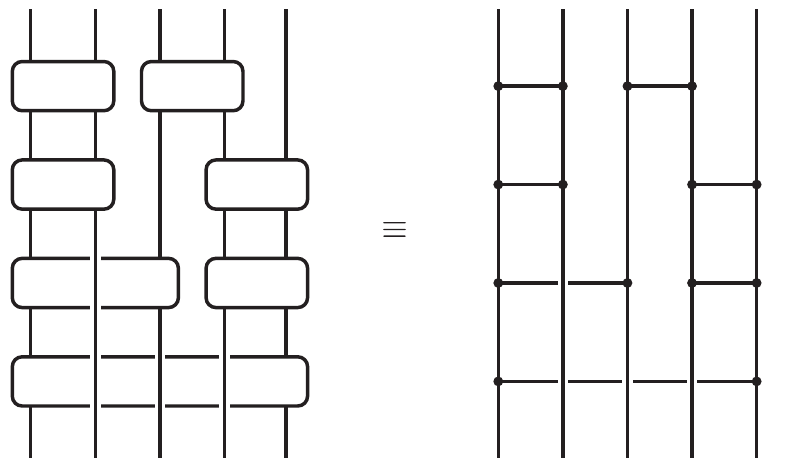}

If the boundary system $\partial \mathcal{M}$ is thought of as having $d$ spacetime ``dimensions'' then the resulting graph with vertex set $X$ is a classical geometrical space having spacetime dimension $d+1$, with the role of the \emph{holographic time} axis being played by the set $\{1,2, \ldots, T\}$.

We follow a slightly different, yet morally equivalent, approach to associating a bulk holographic geometry to a boundary system in this paper, where the holographic time dimension is continuous. We detail this idea in the next section.

\section{Bulk topology and geometry from geodesics in $\textsl{SU}(\mathcal{H})$}\label{sec:btg}
In this section we explain how to associate a bulk topological space to any path $\gamma$ in $\textsl{SU}(\mathcal{H})$ connecting the identity $\mathbb{I}$ to a unitary $U$ acting on the boundary space. 

Let $\gamma$ be a path connecting $\mathbb{I}$ to $U$ in $\textsl{SU}(\mathcal{H})$. As a matrix we express $\gamma$ as a time-ordered product
\begin{equation}
  \gamma \equiv \mathcal{T}e^{-i\int_0^R K(r)\,dr},
\end{equation}
where $K(r) \in \mathcal{B}(\mathcal{H})$ is a possibly time-dependent traceless hermitian operator generating the evolution at $\gamma(r)$. The matrix $K(r)$ may be regarded as a time-dependent Hamiltonian acting on the boundary system. We can express $K(r)$ as a sum of interaction terms acting on the subsystems of $\partial \mathcal{M}$:
\begin{equation}
  K(r) = \sum_{I\subset \{1,2, \ldots, n\}} k_I(r),
\end{equation}
where $k_I(r)$ is an operator acting nontrivially only on the subsystems in the subset $I$. In general, for the metrics we consider here, all possible subsets $I$ can appear, and there are exponentially many (in $n$) interaction terms. In other words, $K(r)$ is generically a strongly interacting quantum spin system. 

We want to associate a topological space to $K(r)$ for each \emph{instantaneous holographic time slice} $r \in [0,R]$. There are many operationally meaningful ways to do this, depending on the physical questions you ask. One way is to interpret $K(r)$ as a \emph{free-particle Hamiltonian} for some possibly very complicated configuration space $\mathcal{X}$ which is built by matching the dispersion relation of the localised excitations of $K(r)$ to that of the free-particle Hamiltonian on $\mathcal{X}$. Another way, one of which we focus on here, is to study the response of high-temperature states $\rho_\beta(r)$, with $\beta$ small, to localised perturbations $A$ and $B$ at different sites: at zero inverse temperature $\beta = 0$ all perturbations on different sites will be completely uncorrelated, however, when $\beta$ is small there are residual correlations between \emph{nearby} sites allowing us to say when two sites are \emph{close}. This approach, while somewhat indirect, has the considerable upside that it immediately leads to a positive-definite metric. Yet another approach is to study the propagation of a localised perturbation $A$ at some site $j$ according to the Schr\"odinger time evolution determined by $K(r)$ and \emph{assuming} a Lieb-Robinson type bound \cite{Lieb:1972a, Nachtergaele:2010a} on the dynamics of $K(r)$:
\begin{equation}
	\|[A(\tau),B]\| \le Ce^{v|\tau|-d(j,k)} \|A\|\|B\|,
\end{equation}
where $C$ is a constant, $v$ is the group velocity, and $B$ is an observable localised at some other site $k$. Such a bound can be used to infer a \emph{pseudo-Riemannian} type structure via a \emph{causality relation} on the set $\{1,2,\ldots, n\}\times \mathbb{R}^{+}$ which can, in turn, be quantified in terms of a \emph{causal set} leading to an embedding in a Lorentz manifold. (Here $\tau$ is the standard time coordinate for the boundary quantum system.) We discuss this idea in the second subsection. These last two proposals may be regarded as a Wick-rotated ``Euclidean approach'' and ``Lorentzian approach'', respectively, to the problem of building bulk holographic spacetimes associated with paths of unitaries.

\subsection{Bulk holographic geometry from thermal correlations}
Suppose that a quantum system of $n$ quantum spins $\{1,2, \ldots, n\}$ with Hamiltonian $K(r)$ is brought into thermal equilibrium at inverse temperature $\beta$: the state of the system is described by the Gibbs ensemble
\begin{equation}
	\rho_\beta(r) \equiv \frac{e^{-\beta K(r)}}{\tr(e^{-\beta K(r)})}.
\end{equation}
Consider the effect of a small perturbation $A\in\mathfrak{su}(\mathcal{H})$ localised at site $j$ (respectively, a small perturbation $B\in\mathfrak{su}(\mathcal{H})$ localised at site $k$): the resulting system state is now
\begin{equation}
	\rho_\beta(r) + \epsilon X \approx \frac{e^{-\beta K(r)+i\epsilon A}}{\tr(e^{-\beta K(r)})}, 
\end{equation}
respectively, 
\begin{equation}
	\rho_\beta(r) + \epsilon Y \approx \frac{e^{-\beta K(r)+i\epsilon B}}{\tr(e^{-\beta K(r)})}.
\end{equation}
(The reason for the factor of $i$ is that elements of $\mathfrak{su}(\mathcal{H})$ are \emph{antihermitian} in this paper.)
Now we ask the question: how \emph{distinguishable} is the perturbed state $\rho_\beta(r) + \epsilon X$ from the state $\rho_\beta(r) + \epsilon Y$? We say that the local perturbation $A$ at site $j$ is \emph{close}, or \emph{adjacent}, to the perturbation $B$ local to site $k$ if the states $\rho_\beta(r) + \epsilon X$ and $\rho_\beta(r) + \epsilon Y$ are \emph{not completely distinguishable}. That this notion corresponds to a topological/geometrical conception of closeness may be argued as follows. If the temperature is very high, i.e., near to the infinite-temperature fixed point $\rho\propto \mathbb{I}$, then all correlations are disordered by thermal fluctuations. The effects of a local perturbation are hence delocalised only in a small surrounding region determined by the high-temperature correlation length, which directly depends on the inverse temperature. Hence, if $\rho_\beta(r) + \epsilon X$ and $\rho_\beta(r) + \epsilon Y$ are independent fluctuations, i.e., they are uncorrelated, we say that $A$ is \emph{far} from $B$, otherwise, they are adjacent. This region, in turn, determines the desired adjacency relation for the sites $j$ and $k$ which, in turn, supplies us with a metric quantity. 

It is a remarkable fact that the quantum informational distinguishability, as measured by the relative entropy $S(\cdot\|\cdot)$, of the states $\rho_\beta(r) + \epsilon X$ and $\rho_\beta(r) + \epsilon Y$ is quantified to $O(\epsilon)$ by the following equation \cite{Beny:2015a, Beny:2013a, Beny:2015b}:
\begin{equation}
	\langle A, B\rangle_{\rho_\beta(r)} \equiv -\frac{\partial^2}{\partial x\partial y} F(x,y)\big|_{x=y=0},
\end{equation}
where $F(x,y)$ is the \emph{free energy}
\begin{equation}
	F(x,y) = -\frac{1}{\beta}\log\left(\tr\left(e^{-\beta K(r)+ixA+iyB}\right)\right).
\end{equation}
This idea has also been exploited in various incarnations by Nozaki, Ryu, and Takayanagi \cite{Nozaki:2012a} to identify metrics for holographic spacetimes and is most directly inspired by the distance quantity exploited by Qi in investigations of the exact holographic mapping \cite{Qi:2013a}. Rather fortuitiously, the quantity $\langle \cdot,\cdot\rangle_{\rho_\beta(r)}$ is a positive definite \emph{inner product} on the space of local operators. Additionally, it is equal to the following two-point thermal correlation function
\begin{equation}
	\langle A, B\rangle_{\rho_\beta(r)} \equiv \frac{1}{\beta}\int_0^\beta \tr\left(\rho_\beta(r) e^{uK(r)}Be^{-uK(r)} A\right)\, du.
\end{equation}
It is this quantity that we employ to determine an adjacency relation between the sites. 

When $\beta$ is infinitesimal the two-point thermal correlation function is given by 
\begin{equation}\label{eq:gapprox}
	\langle A, B\rangle_{\rho_\beta(r)} \approx \frac{1}{2^n}\tr(A B) - \frac{\beta}{2^{n+1}}\tr(A\{K(r),B\}) + O(\beta^2).
\end{equation}
However, we also know \cite{Hastings:2006a, Kliesch:2014a} that the high-temperature two-point correlation functions are exponentially decaying for $\beta$ small: 
\begin{equation}\label{eq:gdecay}
	|\langle A, B\rangle_{\rho_\beta(r)}| \lesssim e^{-\frac{d(j,k)}{\xi(\beta)}}\|A\|\|B\|,
\end{equation}
where, generically, the high-temperature correlation length tends to zero like $\xi(\beta)\propto \beta$ as $\beta\rightarrow 0$. (The exponential decay of high-temperature correlations notably does \emph{not} hold for bosonic systems, and we must resort to other means in this case.) Thus, if $\langle A, B\rangle_{\rho_\beta(r)}$ is nonzero for $\beta$ infinitesimal when $j\not=k$ this means that $d(j,k)$ must be arbitrarily small, i.e., $j$ and $k$ are \emph{adjacent}.

Our task is thus to extract a distance measure, or metric, $d(j,k)$ from $\langle A, B\rangle_{\rho_\beta(r)}$. One direct way of doing this is simply to take a log of Eq.~(\ref{eq:gdecay}), i.e., define
\begin{equation}\label{eq:metricfirst}
	d(j,k) \overset{!}{\equiv}  \sup_{A,B} - \beta \log \frac{|\langle A, B\rangle_{\rho_\beta(r)}|}{\|A\|\|B\|},
\end{equation}
similar to the approach of Qi \cite{Qi:2013a}. Unfortunately, it is not clear if $d(j,k)$ so defined satisfies the triangle inequality $d(j,l)\le d(j,k)+d(k,l)$. We will evade this problem by using Eq.~(\ref{eq:metricfirst}) only to identify an \emph{adjacency relation} between pairs of spins $(j,k)$ and then use this adjacency relation to build a metric. What this means is we first set up the \emph{adjacency matrix}
\begin{equation}
	A_{j,k} = \sup_{A,B} - \beta \log \frac{|\langle A, B\rangle_{\rho_\beta(r)}|}{\|A\|\|B\|}, \quad j\not=k.
\end{equation}
This defines a weighted graph structure $G=(V,E)$ on the vertex set $V=\{1,2,\ldots, n\}$. For any pair of points $j$ and $k$ in $G$ we define the distance between $j$ and $k$ as the length of the shortest path $p= (e_1,e_2, \ldots, e_m)$, where $e_l = (x_l,y_l)$ are edges, between $j$ and $k$. This is guaranteed to obey the triangle inequality. Thus we define the metric $d(j,k)$ according to 
\begin{equation}
	d(j,k) = \inf\left\{\sum_{(x,y)\in p} A_{x,y}\,\middle| \text{$p$ is a path from $j$ to $k$}\right\}.
\end{equation}

The definition of the metric we supply in this subsection is difficult to compute in general. We can build a computable approximation by comparing Eq.~(\ref{eq:gapprox}) expanded to first order and Eq.~(\ref{eq:gdecay}): if $\tr(A\{K(r),B\}) \lesssim e^{-\frac{1}{\beta}}$ for all $A$ and $B$ then $j$ and $k$ are not adjacent. If, however, there are local operators $A$ at $j$ and $B$ at $k$ such that for $\beta$ infinitesimal 
\begin{equation}
	\langle A, B\rangle_{\rho_\beta(r)} \gg e^{-\frac{1}{\beta}},
\end{equation}
then $j$ and $k$ \emph{are} adjacent. Restricting our attention to hamiltonians $K(r)$ comprised of only one- and two-particle interaction terms $k_{j,k}(r)$ (this is the case when $p\rightarrow \infty$) then to first order in $\beta$ this is equivalent to asking if there are traceless operators $A$ at $j$ and $B$ at $k$ such that 
\begin{equation}\label{eq:connectivity}
	\tr(A\{K(r),B\}) \not= 0,
\end{equation}
i.e., $j$ is adjacent to $k$ if the two-particle interaction term $k_{j,k}(r)$ in $K(r)$ is nonzero. Physically this is equivalent to saying that $j$ and $k$ are adjacent if at time $r$ an (infinitesimal) quantum gate was applied coupling $j$ and $k$. In the case where $K$ is comprised of three-particle or higher interactions we need to go to higher orders in $\beta$ to determine a connectivity relation (at first order the condition Eq.~(\ref{eq:connectivity}) misses three-particle interactions, we need to go to $O(\beta^2)$ to see the effect of such terms).

Taking the product of the metric topology determined by $d(\cdot,\cdot)$ for each $r$ gives us our desired bulk topological space $\mathcal{M}$. 

\subsection{Bulk holographic geometry from causal sets}
The method described in the previous subsection, while giving rise to a metric topological space, does not really capture an important aspect of quantum circuits comprised of local gates, namely, their \emph{causal structure}: in every quantum circuit there is a kind of ``light cone'' of information propagation where we can say that qubit $j$ is in the \emph{past} of qubit $k$ if there is a sequence of quantum gates in the circuit connecting $j$ to $k$. Because the geodesics $\gamma$ in $\textsl{SU}(\mathcal{H})$ obtained via the principle of minimal complexity are generated by essentially local gates this strongly suggests we should actually rather associate some kind of discretised \emph{pseudo-Riemannian} manifold to the bulk holographic spacetime. In other words, it is rather more natural to think of $\mathcal{M}$ as a de Sitter-type space \cite{Beny:2013b, Czech:2015a, Czech:2015b}. Equivalently, one should regard the approach of the previous section as the Wick-rotated Euclidean version of the approach described here.

In this subsection we detail an alternative approach to determining a bulk holographic geometry from a path $\gamma$ in $\textsl{SU}(\mathcal{H})$ by associating a \emph{causal set} $X$ \cite{Bombelli:1987a, Brightwell:1991a} to $\gamma$. Causal sets, in turn, are naturally associated to embeddings in pseudo-Riemannian manifolds. 

Before we describe our construction we briefly review the main ideas of causal sets. A \emph{causal set} is a \emph{locally finite partially ordered set} $X$ of events, i.e., a set with order relation $\preceq$ which is \emph{reflexive} (i.e., $x\preceq x$), \emph{transitive} (i.e., $x\preceq y\preceq z$ implies $x\preceq z$), and \emph{noncircular} (i.e., $x\preceq y\preceq x\not=y$ is excluded). To explain what ``locally finite'' means we introduce the idea of an \emph{Alexandroff set} which is a set of the form
\begin{equation}
	[x,y] \equiv \{z\,|\, x\preceq z\preceq y\};
\end{equation}
if every Alexandroff set $[x,y]$, $x,y\in X$, contains a finite number of elements then $X$ is said to be locally finite. A topology $\mathcal{T}$ may be placed on $X$ by using the Alexandroff sets as a base. 

To describe distances in causal sets we introduce the notion of a \emph{chain} $C$ which is a subset of $X$ such that for all pairs $x$ and $y$ in $X$, $x$ and $y$ can be compared via $\preceq$, i.e., either $x\preceq y$ or $y\preceq x$. Thus $C$ is a sequence $x=x_1\preceq x_2 \preceq \cdots \preceq x_s=y$. The distance $d(x,y)$ between $x$ and $y$ is now defined to be the $s-1$, where $x=x_1\preceq x_2 \preceq \cdots \preceq x_s=y$ is a \emph{maximal chain} connecting $x$ to $y$.

To obtain a causal set $X$ from a path $\gamma \equiv \mathcal{T}e^{-i\int_0^T K(r)\,dr}$ we sample points from the Poisson distribution on $\{1,2,\ldots, n\}\times [0,T]$ with density $\varrho$. This gives us, almost surely, a finite set $X$ of points. We then build a causality relation on this set by first choosing a \emph{threshold} $\epsilon$ and then setting $x\preceq y$ if it is possible to send a \emph{detectable signal} from $x = (j,x_0)$ to $y = (k,y_0)$ via the unitary process $\gamma$. To obtain a causal set structure one has to allow for arbitrary fast local interventions via local unitary operations (LU) during the evolution of the unitary process $\gamma$: what this means is that we are allowed to interrupt the evolution $\gamma(t) = \mathcal{T}e^{-i\int_0^t K(r)\,dr}$ at any holographic time $t$, locally adjoin ancillary quantum systems initialised in some pure state $|0\rangle$, and apply an arbitrary product unitary operation of the form $U_1\otimes U_2\otimes \cdots U_n$ on $\mathcal{H}\otimes \mathcal{H}_{\text{anc}}$, where $\mathcal{H}_{\text{anc}}$ is the Hilbert space for the additional ancillary degrees of freedom. Such operations do not allow additional information transfer between the subsystems. We write any evolution from holographic time $t=x_0$ to holographic time $t=y_0$ resulting from such arbitrary local unitary interventions as a \emph{completely positive} (CP) map $\mathcal{E}_{y_0, x_0}$. We now obtain a causal set structure by saying that $x\preceq y$ if there exist operators $A$ and $B$ local to sites $j$ and $k$, respectively, such that (assuming, without loss of generality, that $x_0<y_0$):
\begin{equation}\label{eq:causalconnect}
	\|[  \mathcal{E}_{y_0,x_0}(A) , B]\| > \epsilon\|A\|\|B\|.
\end{equation}

This way of associating causal structures to a path $\gamma$ in $\textsl{SU}(\mathcal{H})$ also gives us a topological space $(X,\mathcal{T})$, this time generated by the Alexandroff sets. The space we obtain is rather different from that obtained in the previous section as a causal set is a pseudo-Riemannian or Lorentzian space. Morally speaking, the topological space obtained in the previous section is the ``Wick rotated'' version of the one obtained here.

As we increase the density of points in $X$ we obtain finer and finer causal sets. It is an intriguing question whether we can obtain a sensible continuum limit
\cite{Rideout:2001a}. 

\section{Complexity, action, and bulk fluctuations}\label{sec:cabf}
The principle of minimal complexity identifies a geodesic $\gamma$ in $\textsl{SU}(\mathcal{H})$ which, in turn, gives rise to a bulk geometry according to the constructions of the previous section. Here we discuss the \emph{fluctuations} of the bulk geometry by introducing an energy functional determining the geodesic $\gamma$ and defining a corresponding partition function for what is presumably a quantum gravity theory.

In Riemannian geometry a geodesic in a manifold $\mathcal{M}$ may be determined by minimising the \emph{energy}
\begin{equation}
	E(\gamma) \equiv \frac12\int_0^{T} \langle \dot{\gamma}, \dot{\gamma}\rangle_{\gamma} \, dt.
\end{equation}
This quantity is minimised precisely on geodesics $\gamma$ achieving the minimum geodesic distance $d(\mathbb{I}, U)$. A \emph{fluctuation} $\gamma'= \gamma + d\gamma$ of a geodesic $\gamma$ therefore should be a \emph{path} in $\textsl{SU}(\mathcal{H})$ which has a near-minimal energy. Since any path in $\textsl{SU}(\mathcal{H})$ gives rise to a bulk geometry, perturbations $\gamma'$ of $\gamma$ can also be interpreted as \emph{fluctuations} in the bulk geometry. If we imagine that the paths $\gamma$ arise from a \emph{quantum system} then it is natural to introduce the partition function
\begin{equation}\label{eq:bgpartfun}
	\mathcal{Z}_B \equiv \int \mathcal{D}\gamma\, e^{-\beta E(\gamma)},
\end{equation}
to model the fluctuations, where $\int\mathcal{D}\gamma$ is the path integral. Clearly, as $\beta\rightarrow \infty$, the integral is dominated by the classical minimiser $\gamma$. Fluctuations $\gamma'$ are determined by the Gibbs distribution. The partition function Eq.~(\ref{eq:bgpartfun}) can be understood as that for a string with target space $\textsl{SU}(\mathcal{H})$ with fixed endpoints at $\mathbb{I}$ and $U$.

What is the structure of a fluctuation? The energy $E(\gamma)$ is sensitive only to the presence of \emph{quantum gates} between pairs of spins but not \emph{which} spins $j$ and $k$ the gate is applied to. Thus it is easy to describe the structure of near-minimal fluctuations of a geodesic: these are equal to $\gamma(t)$ for all $t$ except at one instant $t=t_w$ when a unitary gate $V_{j,k}$ is applied to an arbitrary pair $(j,k)$ followed immediately by its inverse $V^\dag_{j,k}$. Such a geodesic corresponds to a bulk holographic spacetime which is equal to the minimal one except with a ``wormhole'' between $j$ and $k$ at $t=t_w$ which immediately ``evaporates''. Thus the fluctuating bulk geometry determined by the partition function Eq.~(\ref{eq:bgpartfun}) is comprised of spacetimes where wormholes are fluctuating in and out of existence between all pairs $(j,k)$ of points. 

The path integral in Eq.~(\ref{eq:bgpartfun}) is remarkably simple in that it is quadratic in the tangent field $-iK(r)$ and hence the path measure $\mathcal{D}\gamma \,e^{-\beta E(\gamma)}$ may be understood as a Brownian measure on paths in the unitary group $\textsl{SU}(\mathcal{H})$ generated by $2$-local tangent vectors. Precisely these Brownian motions on the unitary group were introduced in \cite{Lashkari:2013a} as a model for black hole dynamics; in the $p\rightarrow \infty$ limit each path $\gamma(t)$ is a solution to the following stochastic differential equation 
\begin{multline}
	d\gamma(t) \propto i\sum_{j\not= k}^n\sum_{\alpha_k=0}^3 \sigma_{j}^{\alpha_j}\otimes \sigma_{k}^{\alpha_k} \gamma(t) \, dB_{\alpha_j\alpha_k}(t)  - \frac12 \gamma(t)\, dt,
\end{multline}
where $dB_{\alpha_j\alpha_k}(t)$ are independent Brownian motions with unit variance per unit time.
What makes the partition function nontrivial is the constraint that the endpoints of the path are exactly $\mathbb{I}$ and $U$, which turns the path integral into an integral over \emph{Brownian bridges} (see, e.g., \cite{Levy:2015a} for details on the Brownian bridge in a unitary group) on $\text{SU}(\mathcal{H})$. In this context, fluctuations in the bulk geometry are interpreted as a very complicated random variable $g \equiv g(U)$ which depends in a rather nonlinear way on the realisation $U$ of the Brownian bridge. 

We end this section with a comment on the relationship of the definition pursued here the recent argument that information complexity equals action in the holographic context \cite{Brown:2015bva, Brown:2015bva2}. The proposal Eq.~(\ref{eq:bgpartfun}) essentially promotes this argument to a \emph{definition}: the action $E(\gamma)$ \emph{is} directly related to the complexity $d(\mathbb{I},U)$ in exactly the same way the energy of a geodesic is related to the geodesic length in Riemannian geometry, i.e., the minima of both quantities coincide.

\section{Boundary perturbations and Jacobi fields}\label{sec:bpjf}
In this section we discuss the effect of a boundary perturbation on the bulk geometry determined by the principle of minimal complexity. We argue that the principle of minimal complexity already determines an equation of motion constraining the structure of the induced bulk fluctuations. This equation of motion could be understood as a kind of generalised Einstein equation. 

The basic idea of this paper is captured by the following diagramme
\begin{center}
	\includegraphics{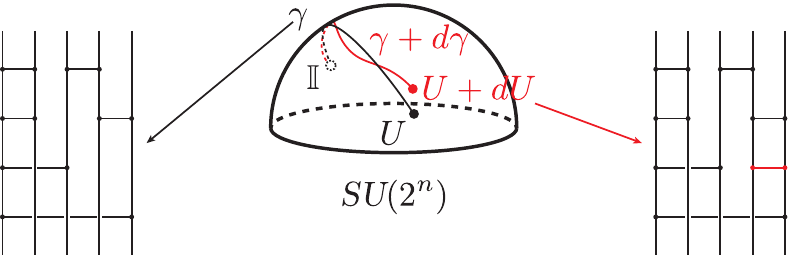}
\end{center}
Suppose the boundary system $\partial \mathcal{M}$ experiences a \emph{fluctuation}. We model this as a perturbation of the unitary $U$, i.e., we study perturbed unitaries $U' = U + dU$. One natural source of such fluctuations arises from the presence of \emph{local external fields} $J$, i.e., we study the unitaries $U(s,J)$ diagonalising the boundary Hamiltonians
\begin{equation}
	H(s,J) \equiv H + s\sum_{j=1}^n\sum_{\alpha = 1}^3 J_\alpha^j \sigma_{j}^\alpha,
\end{equation}
where $J_\alpha^j$ is a collection of $3n$ numbers parametrising an arbitrary inhomogeneous external field and $s$ is an infinitesimal. Knowledge of the ground state $|\Omega(s,J)\rangle$ of a gapped Hamiltonian $H(s,J)$ for all $J$ allows us to calculate the expectation value $\langle\Omega|\sigma^{\boldsymbol{\alpha}}|\Omega\rangle$, for any collection of $\boldsymbol{\alpha}\in \{0,1,2,3\}^{\times n}$ by differentiation with respect to $J$ at $s=0$. The unitary $U(s,J)$ is the \emph{generating function} for $H$. Another natural source of fluctuations comes from unitaries of the form $U(s, M) = e^{-isM}U$, with $M\in\mathcal{B}(\mathcal{H})$ a hermitian operator and $s$ small. The physical justification for such fluctuations comes from interpreting $U$ as the quantum circuit which prepares the boundary system in a low-energy eigenstate of the boundary hamiltonian $H$. A circuit such as $U(s,M) = e^{-isM}U \approx U + dU$ represents the situation where some particles fluctuated into existence after the system was prepared in the low-energy sector.

So long as $\mathbb{I}$ and $U$ are \emph{not} conjugate points we can apply the prescription of the previous section to identify a \emph{family} of geodesics $\gamma(r,s)$ connecting $\mathbb{I}$ to $U(s,J)$ or $U(s,M)$ near to the geodesic $\gamma$ connecting $\mathbb{I}$ to $U$, i.e., we study  first-order corrections
\begin{equation}
	\gamma(r,s) \approx \gamma(r) + s \partial_s\gamma(r,s)|_{s=0}.
\end{equation}
Via the argument of the previous section a shift in $\gamma(r)$ corresponds in a shift $\mathcal{M}\mapsto \mathcal{M} + d\mathcal{M}$ in the bulk holographic spacetime. Since we capture the structure of the bulk holographic spacetime with a (metric) topology, i.e., we observe a shift in the topology $\mathcal{T}$ on the point set $X$. The key point is now that the vector field $\partial_s\gamma(r,s)$ which captures the first-order shift in $\gamma(r)$ is \emph{far from} arbitrary, indeed, it satisfies a remarkable nontrivial equation of motion known as the \emph{Jacobi equation}: 
\begin{multline}
	\partial_r^2 Y = B_p(\partial_r Y + [X,Y],X) + B_p(X,\partial_r Y + [X,Y]) \\ - [B_p(X,X),Y] + [X,\partial_r Y],
\end{multline}
where we've defined $X \equiv (\partial_r\gamma) \gamma^{-1}$ and $Y \equiv (\partial_s\gamma) \gamma^{-1}$ \cite{Arnold:1966a, Arnold:1989a, Wald:1984a}. This is a second-order equation of motion for the fluctuation $Y$.

Since fluctuations in geodesics $\gamma(r)$ directly correspond to fluctuations in bulk geometries the Jacobi equation may be naturally regarded as a kind of ``Einstein equation'' constraining the dynamics of the bulk geometrical fluctuations. The vector field $Y$ capturing the bulk geometrical fluctuation $d\mathcal{M}$ is directly a function of the external boundary field $J_\alpha^j$, allowing us to deduce a precise bulk/boundary correspondence. This observation is the main contribution of this paper. 

For arbitrary local $H$ it is very hard to say anything nontrivial about the structure of $U(J)$, and hence $Y$, so our general conclusions concerning the properties of the fluctuation field $Y$ are consequently limited; only in the context of solvable examples can we say anything more.

\begin{figure}
\includegraphics{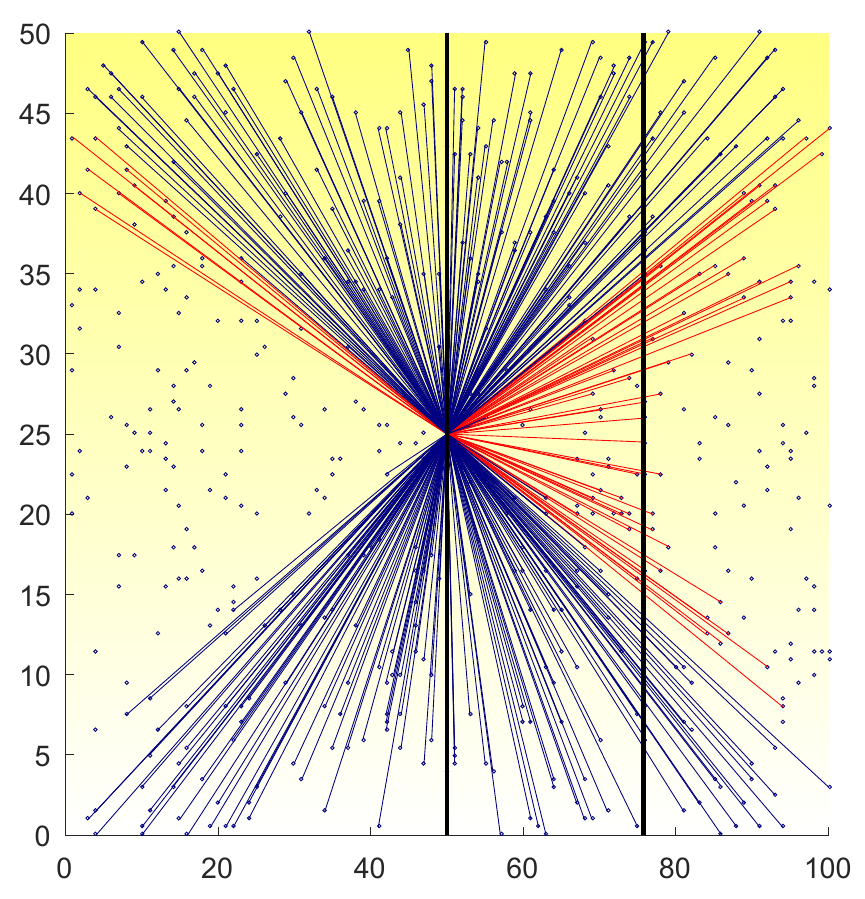}
\caption{Example of the fluctuation in bulk spacetime $\mathcal{M}$ and bulk causal structure due to a fluctuation on the boundary. The boundary quantum system $\partial \mathcal{M}$ is comprised of $n=100$ qubits, and the boundary Hamiltonian is given by the $1D$ nearest-neighbour transverse Ising model $H = \sum_{j=0}^{100} \sigma_j^x\sigma_{j+1}^x + h\sigma_j^z$, with periodic boundary conditions. The $x$ axis is labelled by site number and the $y$ axis is holographic time $r$. The dots represent events in bulk holographic spacetime and have been chosen according to the Poisson distribution. The unitary operator $U$ studied here is $U = e^{-i50 H}$, a quench scenario. We studied the minimal geodesic $\gamma(r) = e^{-irH}$ connecting the identity $\mathbb{I}$ to $U$. The blue lines illustrate causal connections from a reference event at $(j=50,r=25)$ to the Poisson distributed events according to the criteria Eq.~(\ref{eq:causalconnect}). We considered a fluctuation $U'= e^{-i\delta h_{50,75}}U$ which models the addition of a remote entangled pair between the distant sites $50$ and $75$ (the spacetime history of both of the involved sites are illustrated with black lines) at time $r=50$. The bulk holographic spacetime for the new geodesic $\gamma'$ connecting $\mathbb{I}$ to $U'$ was calculated according to the principle of minimal complexity by solving the Jacobi equation and the additional causal connections illustrated in red. One can readily observe the change in spacetime topology induced by the fluctuation, which might be interpreted as the creation of a wormhole between sites $50$ and $75$.}\label{fig:fluctations}
\end{figure}

\section{Examples}\label{sec:examples}
Unfortunately, except for all but the simplest cases, the geodesic $\gamma$ connecting $\mathbb{I}$ to a unitary $U$ is very hard to calculate, especially when $p\not=1$. Nevertheless, much can already be learned from very simple examples. 

\subsection{Example 1: the trivial case; bulk background}
Suppose the boundary system is \emph{trivial}, i.e., the unitary rotating $H$ to its eigenbasis is simply $U = \mathbb{I}$. This would be the case, e.g., for the noninteracting boundary system
\begin{equation}
	H = \sum_{j=1}^n \sigma_j^z.
\end{equation}
In this case $C_p(U) = 0$ for all $p$ and the holographic time direction collapses to a point set. The associated holographic geometry is also trivial: This example corresponds to a set of $n$ completely disconnected bulk universes. The fluctuations are also structureless as all different pairs of sites $j\not=k$ fluctuate indendently, corresponding to spontaneous creation and annihilation of wormholes between all pairs of sites.  

\subsection{Example 2: the trivial case; pairwise perturbations}
Imagine the trivial example experiences a boundary fluctuation where a pair $(i,j)$ of boundary spins is spontaneously entangled: $H \mapsto V_{j,k}^\dag H V_{j,k}$, where $V_{j,k}$ is a near-identity unitary operation entangling spins $j$ and $k$. For example, take $V_{j,k} = e^{-i\epsilon \sigma_j^x\sigma_k^x}$. In this case $H$ fluctuates to 
\begin{equation}
	H' \equiv H + i\epsilon (\sigma_j^y\sigma_k^x + \sigma_j^x\sigma_k^y)
\end{equation}
By construction the unitary $U'$ diagonalising $H'$ is simply $U' = V_{j,k} = \mathbb{I} -i\epsilon \sigma_j^x\sigma_k^x$.

It is straightforward to calculate the new geodesic $\gamma'$ connecting $\mathbb{I}$ to $U'$: it is simply 
\begin{equation}
	\gamma'(r) \equiv e^{-ir\sigma_j^x\sigma_k^x}.
\end{equation}
The causal structure of the fluctuation in the associated bulk geometry may be directly described: sites $j$ and $k$ become causally connected while the remaining sites remain causally disconnected. 

\subsection{Example 3: quench dynamics}
The final example we cover here concerns unitaries of the form $U=e^{i\tau L}$, with $L\in\mathcal{B}(\mathcal{H})$ a local generator. This sort of unitary is natural when studying the dynamics of \emph{quenched systems} where the hamiltonian of the boundary quantum system is suddenly changed from some initial hamiltonian $H$ to a new hamiltonian $L$. Recently it has been argued that such dynamics are dual to Einstein-Rosen bridges supported by localised shock waves \cite{Roberts:2015a}. The boundary system experiences a rotation according to $L$. In this particular case it is rather easy to solve the Euler-Arnol'd equation (as long as $\mathbb{I}$ and $U$ are not conjugate points), namely, we find the geodesic
\begin{equation}
	\gamma(r) \equiv e^{ir L}, \quad r \in [0,\tau],
\end{equation}
that is, the vector field $-iK(r)$ is constant and simply equal to $L$. 

Consider now a fluctuation of the form $U' = e^{isM}U$, with $M$ local to a pair $(j,k)$ of sites, representing a nonlocal entangled pair of particles fluctuating into existence at sites $j$ and $k$ just after the quench. In this rather general case we can actually completely solve the Jacobi equation to yield the (constant) vector field $Y$:
\begin{equation}
	-iY(r) = \int_0^\infty \frac{\mathbb{I}}{U + u\mathbb{I}} M\frac{U}{U + u\mathbb{I}}\,du.
\end{equation}
(Although not manifestly hermitian this expression does indeed lead to a hermitian operator which can be confirmed by directly evaluating the integral.)

We have illustrated the application of this formula in Fig.~\ref{fig:fluctations} where we've calculated the causal structure of the bulk spacetime geometry according to a fluctuation of a boundary quantum system given by the transverse Ising model.

\section{Conclusions and outlook}\label{sec:conclusions}
In this paper we have discussed how, motivated by quantum information considerations, one might associate a bulk holographic spacetime, as a topological space, with an \emph{arbitrary} boundary quantum system. This approach, exploiting the principle of minimal complexity, was directly informed by the recent arguments of Maldacena, Ryu, Takayanagi, van Raamsdonk, Swingle, and Susskind, and others. We introduced two ways to build bulk holographic topological spaces from paths in the unitary group which are morally ``Wick rotated'' versions of each other. Building on this observation we then argued that the principle of minimal complexity supplies us with much more, namely, a quantum model for fluctuations of the bulk holographic spacetime via Brownian bridges on the unitary group. The connection between boundary fluctuations and bulk fluctuations is also similarly determined via minimal complexity considerations: we derived an equation of motion constraining the holographic fluctuations due to low-energy perturbations of the boundary theory. Finally, we illustrated these ideas in the context of several simple examples. 

We have just scratched the surface of these ideas and an enormous number of fascinating questions remain to be explored. A partial list includes: 
\begin{enumerate}
\item The calculations we carried out in this paper are almost exclusively for the case $p=1$ for the metric on $\textsl{SU}(\mathcal{H})$. It is an intriguing question whether any quantitative results can be obtained for the more pertinent limit $p\rightarrow \infty$. At least the Euler-Arnol'd equation of motion can be written out and solved for small $r$. Also, the Jacobi equation is, in principle, solvable for such limits. 
\item The principle of minimal complexity is strongly reminiscent of the principle of least action; indeed, we promoted it per definition to a least action principle to obtain a model for the bulk holographic spacetime fluctuations. This is by far not the first time such ideas have been proposed; indeed we learnt of very similar ideas long ago from Andre Soklakov \cite{Soklakov:2002a}. It is an intriguing question whether there is indeed a deeper connection here between the minimal complexity principle and Kolmogorov complexity, and similarly, between fluctuations and Solomonoff induction. 
\item Should we give in to temptation and interpret the partition function Eq.~(\ref{eq:bgpartfun}) as a quantum gravity theory? Does this theory enjoy any kind of diffeomorphism invariance? As it is a theory of strings in a ridiculously high-dimensional space (namely, the manifold $\textsl{SU}(\mathcal{H})$) can it be related to string theory proper, or is this a mirage?
\item Our boundary quantum system is completely arbitrary, however, it is vitally important to study the continuum limit. This can indeed be done following the method introduced in \cite{Continuouslimits}. The resulting bulk spacetime for CFTs should then converge to AdS. 
\item Tensor networks did not play a prominent role here, but they should emerge as (almost) geodesics. In particular, the perfect tensor model of \cite{Pastawski:2015a} and the EHM of Qi \cite{Qi:2013a}, are most natural candidates. Fluctuations around these cases should be particularly relevant for AdS/CFT dualities. 
\item We only looked at one example in any depth, namely, the transverse Ising model. It would be very interesting to look deeper at more examples, including, more general quantum lattice models and models of black holes, shockwaves, and beyond. 
\end{enumerate}

\acknowledgments
We are grateful for helpful conversations with many people, including, Cedric Beny, Courtney Brell, Seth Lloyd, Brian Swingle, Frank Verstraete,  and Guifre Vidal, amongst many others. This work was supported by the ERC grants QFTCMPS and SIQS, and by the cluster of excellence EXC201 Quantum Engineering and SpaceTime Research.

\end{document}